\documentclass[12pt,a4paper]{article}

\usepackage[left=3cm,top=3cm,bottom=3cm,right=3cm]{geometry}

\usepackage{graphicx}
\usepackage{amssymb}
\usepackage{amsthm}
\usepackage{amsmath}
\usepackage{amsfonts}
\usepackage{algorithm, algorithmic}
\usepackage{color}

\usepackage{cite}

\usepackage{lineno}

\usepackage{enumitem}

\newtheorem{thm}{Theorem}[section]
\newtheorem{cor}[thm]{Corollary}

\theoremstyle{definition}

\theoremstyle{remark}

\numberwithin{equation}{section}

\newcommand{\thmref}[1]{Theorem~\ref{#1}}

\makeatletter
\renewcommand*\env@matrix[1][c]{\hskip -\arraycolsep
  \let\@ifnextchar\new@ifnextchar
  \array{*\c@MaxMatrixCols #1}}
\makeatother

\begin{document}


\title{Idempotent permutations}
\author{A. Emre CETIN}

\maketitle

\begin{abstract}


Together with a characteristic function, idempotent permutations uniquely determine idempotent maps, as well as their linearly ordered arrangement simultaneously. Furthermore, in-place linear time transformations are possible between them. Hence, they may be important for succinct data structures, information storing, sorting and searching. 

In this study, their combinatorial interpretation is given and their application on sorting is examined. Given an array of $n$ integer keys each in $[1,n]$, if it is allowed to modify the keys in the range $[-n,n]$, idempotent permutations make it possible to obtain linearly ordered arrangement of the keys in $\mathcal{O}(n)$ time using only $4 \log n$ bits, setting the theoretical lower bound of time and space complexity of sorting. If it is not allowed to modify the keys out of the range $[1,n]$, then  $n+4 \log n$ bits are required where $n$ of them is used to tag some of the keys.



\end{abstract}




\section{Introduction}\label{sec:intro_abs}

Let $[n]$ denotes the set $\lbrace 1,2,\ldots,n \rbrace$. Given a permutation $\sigma$ of $[n]$, its inverse $\sigma^-$ can be found in-place in $\mathcal{O}(n)$ time. The only way to compute $\sigma^-$ is to somehow tag each inverted element while following the cyclic structure of $\sigma$. This is not a problem if additional $n$ bits are available. On the other hand, if the memory resources are critical or the size $n$ of the permutation is too large, it is possible to tag an element by making negative when it is inverted~\cite[p. 176]{knuth:vol1}. After all the elements are inverted, they can be restored by correcting the signs.

 A map $\iota(1),\iota(2),\ldots,\iota(n)$ is called idempotent if and only if $\iota(\iota(x))=\iota(x)$, for all $x \in [n]$~\cite[p. 91]{comtet}. If the cardinality of the image $\iota([n])$ is $k$, then $\iota$ has $k$ distinct and fixed elements $y \in [n]$ such that $\iota(y)=y$ and the remaining $(n-k)$ elements $z \in [n]$ are idle and equivalent to those $k$ fixed elements such that $\iota(z)=y$. It should be noted that every map from $[n]$ into itself can be rearranged into an idempotent map.

Every idempotent map can be represented with a unique idempotent permutation $\pi(1),\pi(2),\ldots,\pi(n)$, which is indeed a regular permutation of $[n]$, but its $1 \le k \le n$ elements, including $1$, are fixed and in increasing order with respect to each other. A characteristic function can be defined {\em explicitly} to tag the fixed elements. On the other hand, it may be possible to tag them by making negative to improve space complexity from algorithm point of view. In such a case, it is said that the characteristic function is {\em implicitly} defined in $\pi$. An example to an idempotent permutation of degree $k=5$ is given below in two line notation, where the characteristic function is implicitly defined.
\begin{equation}\label{eqn:0}
\pi=
\begin{pmatrix}[r]
1 &  2 & 3 & 4 &  5 & 6 &  7 & 8 &  9 & 10 \\
3 & -1 & 6 & 8 & -4 & 7 & -5 & -9 & -10 & 2 
\end{pmatrix}
\end{equation}

Either defined explicitly or implicitly, the characteristic function and the idempotent permutation $\pi$ uniquely determine simultaneously the idempotent map $\iota$, and its linearly ordered arrangement. The possible transformations covered in this study are as follows.

\begin{enumerate}[label=(\roman{*}).]

\item Every map from $[n]$ into itself can be unstably rearranged in-place into an idempotent map in $\mathcal{O}(n)$ time using $2\log n$ bits. If stability is important, it is possible to determine the regular permutation $\sigma$ of $[n]$ in $\mathcal{O}(n)$ time using $2 \log n$ bits, which can rearrange the map in-place into the corresponding idempotent map stably in further $\mathcal{O}(n)$ time using $4 \log n$ bits. However, this requires additional $n \log n$ bits to store $\sigma$.

\item Any algorithm that inverts a regular permutation $\sigma$ of $[n]$ can be used to invert an idempotent permutation $\pi$. However, if the characteristic function is implicitly defined in $\pi$, in-place inversion algorithms~\cite[p. 176]{knuth:vol1} can not be used due to the fact that it is not possible to tag inverted elements by making negative since fixed elements of $\pi$ are already negative. Thus, it is mandatory to use additional $n$ bits to tag each inverted element. On the other hand, if the characteristic function is explicitly defined, inverted elements can be tagged by making negative and in-place inversion algorithms can be used.

\item Given the idempotent map $\iota$, defining the characteristic function implicitly, the corresponding idempotent permutation $\pi$ can be determined in-place of $\iota$ in $\mathcal{O}(n)$ time using $ \log n$ bits.

\item Given the idempotent permutation $\pi$ of degree $k$, the corresponding idempotent map $\iota$ can be determined in-place of $\pi$ in $\mathcal{O}(k n)$ time using $4 \log n$ bits.

\item Given the inverse $\pi^-$ of the idempotent permutation, the idempotent map $\iota$ can be determined from $\pi^-$ in a separate output array in $\mathcal{O}(n)$ time using $2\log n$ bits.

\item Given the inverse $\pi^-$ of the idempotent permutation, the linearly ordered arrangement of $\iota$ can be determined in-place of $\pi^-$ in $\mathcal{O}(n)$ time using $\log n$ bits. It is important to note that it is not possible to recover back the original idempotent permutation $\pi$ or its inverse $\pi^-$, unless a separate output array is used for determining the linearly ordered arrangement of $\iota$.

 
\item  Although additional $n$ bits are required to invert $\pi$ in-place, if only the linearly ordered arrangement of $\iota$ is required at the end, using $4 \log n$ bits, it is possible to in-place invert fixed elements of $\pi$ while unfixed, hence idle elements are in-place permuted, resulting a sequence in $\mathcal{O}(n)$ time from where the linearly ordered arrangement of $\iota$ can be obtained {\em in-place} in $\mathcal{O}(n)$ further time. In-place inverting fixed elements while permuting idle elements is an operation that can be defined on idempotent permutations having a combinatorial interpretation, and will be called {\em associative permuting} since it is a combination of permuting and inverting.

\end{enumerate}

One of the important consequences of the above transformations is that, given a map $f$ from $[n]$ into itself,
\begin{enumerate}[label=(\roman{*}).]
\item the map $f$ can be rearranged unstably into an idempotent map $\iota$ in $\mathcal{O}(n)$ time using $2\log n$ bits,
\item defining the characteristic function implicitly, the idempotent permutation $\pi$ can be determined in-place of $\iota$ in $\mathcal{O}(n)$ time using $\log n$ bits,
\item the linearly ordered arrangement of $\iota$ (and hence $f$) can be obtained in-place of $\pi$ in $\mathcal{O}(n)$ time using $4 \log n$ bits with associative permuting,
\end{enumerate}
resulting in an algorithm setting the theoretical lower bound of time and space complexity of sorting $n$ integer keys each in $[1,n]$, whereas distribution counting sort, address calculation sort and bucket sort family of algorithms require at least additional $n \log n$ bits~\cite{Seward,Feurzig,Isaac,Tarter,Flores,Jones,Gupta,Suraweera,mahmoud:2000,Cormen}
.

The organization of the study is as follows. First the idempotent maps will be reanalyzed in Section~\ref{sec:I} with a different combinatorial interpretation, which will be important for defining the one-to-one correspondence between them and the idempotent permutations. Then idempotent permutations and their relation with idempotent maps as well as their linearly ordered arrangement will be analyzed in Section~\ref{sec:IP}. Afterwards, three different sorting algorithms will be examined in Section~\ref{sec:application}, and finally the conclusion will follow.



\section{Idempotent maps}\label{sec:I}

Let $[n]$ denotes the set $\lbrace 1,2,\ldots,n \rbrace$ and $F(n)$ be the set of all maps from $[n]$ into itself. The cardinality of $F(n)$ is $n^n$. Every $f \in F(n)$ can be represented with a sequence $f(1),f(2),\ldots,f(n)$, and it is idempotent if and only if $f(f(x))=f(x)$, for all $x \in [n]$.

Let $I(n)$ be the set of idempotent maps of $F(n)$, in which every $\iota \in I(n)$ satisfy
\begin{equation} \label{eqn:8}
\iota(\iota(x))=\iota(x) \text{ for all } x \in [n]
\end{equation}
and $I(n,k)$ be the subset of those for which $k$ is the cardinality of the image $\iota([n])$.

\begin{thm}\label{thm:I}
For $1 \le k \le n$
\begin{equation}
\vert I(n) \vert = \sum^n_{k=1} \vert I(n,k) \vert = \sum^n_{k=1} \binom{n}{k} k^{(n-k)}
\end{equation}
\end{thm}

\begin{proof} \label{proof:2} Every $\iota \in I(n,k)$ is a result of following tasks:
\begin{enumerate}[label=(\arabic{*}).]
\item Select $k$ {\em elements} $A=\lbrace a_1,a_2,\ldots,a_k \rbrace$ from $[n]$ in $\binom{n}{k}$ ways, and {\em fix} them in $\iota$ such that $\iota(a_i)=a_i$, for $i=1,2,\ldots,k$.

\item Since $k$ elements are fixed, $(n-k)$ {\em idle} elements remain. Select {\em nonnegative} multiplicity $(c'_i-1)$ of each {\em idle} $a_i$ in $\iota$, for $i=1,2,\ldots,k$, such that $c'_1+c'_2+\ldots + c'_k=n-k$, in $\binom{n-1}{k-1}$ ways (the number of {\em nonnegative} integral solutions). This is equivalent to selecting strictly increasing $(k-1)$ elements of $C=\lbrace 1,c_2,\ldots,c_k, n+1 \rbrace$ from $[n]$ in $\binom{n-1}{k-1}$ ways, where the difference sequence $c'_i = c_{i+1} - c_{i}$ is of length $k$ and $\sum^k_{i=1} c'_i=n$, since the fist and last elements are $1$ and $(n+1)$, respectively.

\item Arrange $(n-k)$ idle elements of $\iota$ in $(n-k)$ empty locations in $\frac{(n-k)!}{(c'_1-1)!(c'_2-1)!\ldots (c'_k-1)!}$ ways for the particular $C$ selected in the previous step.
\end{enumerate} 
The first $2$ tasks are independent. Hence, if the arrangement of the idle elements of $\iota$ is ignored, the cardinality of $I(n,k)$ is
\begin{equation} \label{eqn:0_0}
\sum^n_{k=1} \binom{n-1}{k-1} \binom{n}{k}
\end{equation}
For all $\binom{n-1}{k-1}$ possible $C$'s, $(n-k)$ idle elements of $\iota$ can be arranged in $$\sum_{c'_1-1+c'_2-1\ldots +c'_k-1=n-k} \frac{(n-k)!}{(c'_1-1)!(c'_2-1)!\ldots (c'_k-1)!} = k^{(n-k)}$$ different ways. Since the number of nonnegative integral solutions of the above summation is $\binom{n-1}{k-1}$, the cardinality of $I(n,k)$ is obtained by multiplying Eqn.~\ref{eqn:0_0} with $k^{(n-k)}$ and dividing by $\binom{n-1}{k-1}$, resulting in $\binom{n}{k} k^{(n-k)}$ and that of $I(n)$ by $\sum^n_{k=1} \binom{n}{k} k^{(n-k)}$.
\end{proof}

\begin{cor}\label{cor:A_B_I}
Let $B=\complement A$. An idempotent map $\iota \in I(n,k)$ divides its domain $[n]$ into $2$ disjoint sets $A$ and $B$ such that, $a \in A$ if and only if $\iota(a)=a$, and there exists an $a \in A$ for every $b\in B$ such that $\iota(b)=a$. This implies together with Eqn.~\ref{eqn:8} that every $\iota \in I(n,k)$ has $k$ distinct and fixed elements $a_1,a_2,\ldots,a_k$, such that $\iota(a_i)=a_i$, for $i=1,2,\ldots,k$.
\end{cor}

\begin{cor}\label{cor:[a_i]}
In addition to dividing its domain $[n]$ into $2$ disjoint sets $A$ and $B$, the idempotent map $\iota \in I(n,k)$ partitions its {\em domain} $[n]$ into $k$ equivalence {\em index} classes,\begin{equation}\label{equiv_class_iota}
[a_i]_\iota := \lbrace x \in [n]  \; \vert \;  \iota(x) = a_i \rbrace \text{ for } i=1,2,\ldots,k
\end{equation}
with cardinalities $c'_1,c'_2, \ldots, c'_k$, respectively. The subscript $\iota$ of $[a_i]$ does not denote a relation. Rather, it denotes that the equivalence {\em index} class $[a_i]$ belongs to $\iota$, since the idempotent permutations will have equivalence index classes which will be denoted by $[a_i]_{\pi}$ in the following section.

For each canonical representative fixed index $a_i \in [a_i]_\iota$ of the fixed element $a_i$, there exist $(c'_i-1)$ idle indices $y \in [a_i]_\iota$ of the idle elements of $\iota$ equivalent to fixed element $a_i$ such that $\iota(y)=a_i$. On the other hand, only the canonical representative fixed index $a_i$ satisfy the equality $\iota(x)=x$. Hence, the equality $\iota(x)=x$ is the choice function for the canonical representative index $a_i$ of each equivalence {\em index} class.

\end{cor}

\begin{thm} \label{thm:F_2_I_ip}
There is an $\mathcal{O}(n)$ time algorithm that rearranges $f \in F(n)$ in-place into $\iota \in I(n)$ unstably using $2\log n$ bits of additional space.
\end{thm}

\begin{proof}
Let $f \in F(n)$ be a map with $k$ distinct elements $a_1,a_2,\ldots,a_k$, with multiplicities $c'_1,c'_2,\ldots,c'_k$. When $f$ is rearranged into an idempotent map, one of each of its $k$ distinct elements will be fixed such that $f(a_i)=a_i$. This is merely processing $f$ for $i=1,2,\ldots,n$, and if $f(f(i)) \ne f(i)$, exchanging $f(i)$ with $f(f(i))$ and continuing with the new element that came to $f(i)$ as in the following algorithm. The algorithm starts with $i \leftarrow 1$ and assumes $D[1\ldots n]$ is the array storing $f$ which will be rearranged in-place into $\iota$ in $\mathcal{O}(n)$ time using $2\log n$ bits.  
\begin{enumerate}[label=(\roman{*}).]
\item If $i > n$, then terminate. Otherwise, if $D[D_i]=D_i$, increase $i$ by one and repeat this step; otherwise, exchange $D_i$ with $D[D_i]$ and repeat this step.
\end{enumerate}
At the end, the map $f \in F(n)$ stored in the array $D[1 \ldots n]$ becomes $\iota \in I(n)$.
\end{proof}

\begin{cor}\label{rearranging_sigma}
Given a map $f \in F(n)$ and a permutation $\sigma(1),\sigma(2),\ldots,\sigma(n)$ of $[n]$, rearranging $f$ in-place according to $\sigma$ such that $f(\sigma(1)),f(\sigma(2)),\ldots, f(\sigma(n))$, is equivalent to performing $n$ simultaneous assignments,
\begin{equation}
f(i) \leftarrow f(\sigma(i)) \quad for \quad i= 1,2,\ldots ,n
\end{equation}
Each $\sigma(i)$ describes the element $f(\sigma(i))$ that should be moved into the place of $f(i)$ while following the cyclic structure of $\sigma$. If $\sigma$ is given in a separate array and it is allowed to modify the given, $f$ can be rearranged in-place according to $\sigma$ very efficiently in $\mathcal{O}(n)$ time using $4\log n$ bits~\cite[ex. 5.2-10]{knuth:vol3}. First $f(1)$ is sent to $j \leftarrow \sigma(1)$ by exchanging $f(1)$ with $f(j)$ which brings $f(j)$ to its final position $f(1)$. If $\sigma(1)$ is exchanged with $\sigma(j)$ as well, making $\sigma(j)=j$, this equality can be used to keep track of the elements of $f$ that have already been rearranged. Then the element moved to $f(j)$ is sent to its final position $j \leftarrow \sigma(1)$. This process continues until $\sigma(1)=1$ which means the actual cycle is permuted. Then the iterator $i$ is increased to continue with $f(2)$ and a new cycle is started if $\sigma(2)\ne 2$. Otherwise, the iterator is increased to continue with $f(3)$. At the end, when all the cycles of $\sigma$ are permuted, all the elements of $f$ arrive to their final position and the association $\sigma(i) = i$ is constructed for every element of $\sigma$, i.e., $\sigma$ becomes $\lbrace 1,2,\ldots,n \rbrace$. Hence, if $D[1\ldots n]$ is the array storing $f(1),f(2),\ldots,f(n)$, it stores $f(\sigma(1)),f(\sigma(2)),\ldots, f(\sigma(n))$ after the rearrangement. 
\end{cor}

\begin{cor}\label{rearranging_sigma^-}
On the contrary, permuting $f \in F(n)$ in-place according to $\sigma^-$ is equivalent to performing $n$ simultaneous assignments~\cite{Fich},
\begin{equation}
f(\sigma^-(i)) \leftarrow f(i) \quad for \quad i= 1,2,\ldots ,n
\end{equation}
Each $\sigma^-(i)$ describes where to move $f(i)$ while following the cyclic structure of $\sigma^-$. If $\sigma^-$ is given in a separate array and it is allowed to modify the given, $f$ can be permuted in-place according to $\sigma^-$ very efficiently in $\mathcal{O}(n)$ time using $3\log n$ bits. First $f(1)$ is sent to its final position $j \leftarrow \sigma^-(1)$ by exchanging $f(1)$ with $f(j)$. If $\sigma^-(1)$ is exchanged with $\sigma^-(j)$ as well, making $\sigma^-(j)=j$, this equality can be used to keep track of the elements of $f$ that have already been permuted. Then the new element that came to $f(1)$ is sent to its final position $j \leftarrow \sigma^-(1)$. This process continues until $\sigma^-(1)=1$ which means the actual cycle is permuted. Then the iterator $i$ is increased to continue with $f(2)$ and a new cycle is started if $\sigma^-(2)\ne 2$. Otherwise, the iterator is increased to continue with $f(3)$. At the end, when all the cycles of $\sigma^-$ are permuted, all the elements of $f$ arrive to their final position and the association $\sigma^-(i) = i$ is constructed for every element of $\sigma^-$, i.e., $\sigma^-$ becomes $\lbrace 1,2,\ldots,n \rbrace$.
\end{cor}

\begin{thm} \label{thm:F_2_I_sigma}
There exists a permutation $\sigma$ of $[n]$ that rearranges $f \in F(n)$ stably into $\iota \in I(n)$, i.e., $\iota(i) = f(\sigma(i))$, for $i =1,2,\ldots,n$, and can be determined in $\mathcal{O}(n)$ time using $2\log n$ bits.
\end{thm}
\begin{proof}
Let $D[1\ldots n]$ be the array storing $f$ and $E[1 \ldots n]$ be the output array which will store $\sigma$. Following algorithm computes $\sigma$ in $E[1\ldots n]$ in $\mathcal{O}(n)$ time using $2\log n$ bits.
\begin{enumerate}[label=(\roman{*}).]
\item Initialize $i,j \leftarrow 1$, and set $E[1]$ through $E[n]$ to zero.
\item If $j > n$, then terminate. Otherwise, if $E[j] \ne 0$, then increase $j$ and repeat this step; otherwise, continue with next step.
\item If $i > n$, then terminate. Otherwise, if $E[D_i] = 0$, then set $E[D_i] \leftarrow i$, increase $i$ by one and goto previous step; otherwise, set $E[j] \leftarrow i$, increase $i$ and $j$ by one and goto previous step;
\end{enumerate}
At the end, the sequence stored in $E[1 \ldots n]$ becomes $\sigma$ that can rearrange $f$ in-place into $\iota$ stably using $4\log n$ bits (Corollary~\ref{rearranging_sigma}).
\end{proof}


\section{Idempotent permutations} \label{sec:IP}

An idempotent permutation $\pi(1),\pi(2),\ldots,\pi(n)$ of degree $k$ is a permutation of $[n]$, with $1 \le k \le n$ elements fixed and in increasing order with respect to each other. The first fixed element of $\pi$ is always $1$. The set of all idempotent permutations will be denoted by $IP(n)$, whereas subset of those of degree $k$ will be denoted by $IP(n,k)$.

\begin{thm}\label{thm:IP}
For $1 \le k \le n$
\begin{equation}
\vert IP(n) \vert = \sum^n_{k=1} \vert IP(n,k) \vert = \sum^n_{k=1} \binom{n}{k} k^{(n-k)}
\end{equation}
\end{thm}
\begin{proof}
Although always will be generated from an idempotent map, an idempotent permutation $\pi \in IP(n,k)$ can be generated from $[n]$, as a result of following tasks:
\begin{enumerate}[label=(\arabic{*}).]
\item Select $k$ {\em indices} $A=\lbrace a_1,a_2,\ldots,a_k \rbrace$ from $[n]$ in $\binom{n}{k}$ ways.
\item Select strictly increasing $(k-1)$ {\em elements} of $C=\lbrace 1,c_2,\ldots,c_k, n+1 \rbrace$ from $[n]$, in $\binom{n-1}{k-1}$ ways, and {\em fix} $1,c_2,\ldots,c_k$ in $\pi$ such that $\pi(a_i)=c_i$, for $i=1,2,\ldots,k$.
\item Arrange the remaining unfixed, hence {\em idle} $(n-k)$ {\em elements} of $\pi$ ($\complement{C}$) into $(n-k)$ empty locations of $\pi$ ($\complement{A}$). Will be explained next.
\end{enumerate}

The first $2$ tasks are independent. Hence, if the arrangement of the idle elements of $\pi$ is ignored, the cardinality of $IP(n,k)$ is
\begin{equation} \label{eqn:1}
\sum^n_{k=1} \binom{n-1}{k-1} \binom{n}{k}
\end{equation}

The difference sequence $c'_i = c_{i+1} - c_{i}$ is of length $k$ and $\sum^k_{i=1} c'_i=n$, since the fist and last elements are $1$ and $(n+1)$, respectively. Hence, $C$ partitions $\pi$ into $k$ equivalence classes
\begin{equation}\label{eqn:IIP_ec}
[c_i]_\pi := \lbrace c_i, c_i+1, \ldots, c_{i+1}-1 \rbrace \text{ for } i=1,2,\ldots,k
\end{equation}
with cardinalities $c'_1,c'_2,\ldots,c'_k$, respectively. For each canonical representative fixed element $c_i \in [c_i]_\pi$, there exist $(c'_i-1)$ idle elements $ c_i+1, \ldots, c_{i+1}-1$ {\em somewhere} in $\pi$ equivalent to $c_i$. Hence, $(c'_i-1)$ idle elements can be arranged in $\frac{(n-k)!}{(c'_1-1)!(c'_2-1)!\ldots (c'_k-1)!}$ different ways for a particular $C$. For all $\binom{n-1}{k-1}$ possible $C$'s, they can be arranged in
\begin{equation}
\sum_{c'_1-1+c'_2-1\ldots +c'_k-1=n-k} \frac{(n-k)!}{(c'_1-1)!(c'_2-1)!\ldots (c'_k-1)!} =k^{(n-k)}
\end{equation}
different ways. Since the number of nonnegative integral solutions of the above summation is $\binom{n-1}{k-1}$, the cardinality of $IP(n,k)$ can be obtained by multiplying Eqn.\ref{eqn:1} with $k^{(n-k)}$ and dividing by $\binom{n-1}{k-1}$, resulting in $\binom{n}{k} k^{(n-k)}$, and that of $IP(n)$ by $\sum^n_{k=1} \binom{n}{k} k^{(n-k)}$.
\end{proof}

\begin{cor}\label{cor:IIP_ec2}
The difference sequence $c'_i = c_{i+1} - c_{i}$ implies $c_{i+1}=c_i + c'_i$. Hence, $k$ equivalence classes  of $\pi$ (Eqn.~\ref{eqn:IIP_ec}) can be represented equivalently by
\begin{equation} \label{eqn:IIP_ec2}
[c_i]_\pi := \lbrace  c_i, c_i+1, \ldots, c_i+c'_i-1  \rbrace \text{ for } i=1,2,\ldots,k
\end{equation}
\end{cor}

Let $\pi \in IP(n,k)$ of degree $k \ge 2$. Using $A$ and $C$, $\pi$ can be represented in two line notation by
\begin{equation*}
\begin{pmatrix}[c]
1      & 2      & \ldots        & a_1-1        &  a_1 & a_1+1        & \ldots & a_2  & \ldots &  a_k & \ldots  &  n \\
\pi(1) & \pi(2) & \ldots        & \pi(a_1-1)    & c_1  & \pi(a_1+1)  & \ldots & c_2  & \ldots &  c_k & \ldots  &  \pi(n) 
\end{pmatrix}
\end{equation*}

\begin{cor}\label{cor:ai_pi}
$C$ does not only partition $\pi$ into $k$ equivalence classes, but also partitions its domain $[n]$ into $k$ equivalence index classes
\begin{equation}\label{eqn:IIP_ec_index}
[a_i]_\pi := \lbrace x \in [n]  \; \vert \;  \pi(a_i) \le \pi(x) < \pi(a_{i+1}) \rbrace \text{ for } i=1,2,\ldots,k
\end{equation}
with cardinalities $c'_1,c'_2,\ldots,c'_k$, respectively. For each canonical representative fixed index $a_i \in [a_i]_\pi$ of fixed element $c_i$, there exist $(c'_i-1)$ idle indices $y \in [a_i]_\pi$ of the idle elements of $\pi$ equivalent to $c_i$ such that $c_i < \pi(y) < c_{i+1}$ which is same with $\pi(a_i) < \pi(y) < \pi(a_{i+1})$.
\end{cor}

There is not a native choice function available for the canonical representative fixed element $c_i$ of $[c_i]_\pi$. Hence a characteristic function is required. Let $\varphi : [n] \mapsto \lbrace 0,1 \rbrace$ be the characteristic function indicating the canonical representative fixed elements. This function indeed indicates the membership of an element in $A$ of $[n]$, as well as equivalently the membership of an element in $C$ of $\pi$, and can be defined by $\varphi(x)=1$ if $x \in A$, hence $\pi(x) \in C$, and $\varphi(x)=0$ otherwise, for all $x \in [n]$. When $\varphi$ is defined implicitly in $\pi$ making fixed elements negative, its definition becomes $\pi(x) < 0$ if $x \in A$, hence $\pi(x) \in C$, and $\pi(x) > 0$ otherwise, for all $x \in [n]$.

\subsection{Inverse of an idempotent permutation}\label{sec:inverse_IP}

Let $\pi \in IP(n,k)$ be an idempotent permutation for which $A = \lbrace a_1,a_2, \ldots,a_k \rbrace$ and $ C = \lbrace 1,c_2,\ldots,c_k, n+1 \rbrace$. If $\pi$ is inverted in-place, the resulting $\pi^-$ would be as follows in two line notation.
\begin{equation*}
\pi^-=
\begin{pmatrix}[c]
c_1   & 2       & \ldots   & c_2-1       & c_2   & c_2+1        & \ldots &  c_k  & \ldots & n \\
a_1  & \pi^-(c_1+1)  & \ldots   & \pi^-(c_2-1)   & a_2  & \pi^-(c_2+1)   & \ldots &   a_k & \ldots & \pi^-(n)
\end{pmatrix}
\end{equation*}

\begin{cor}\label{cor:inverse_I}
It is immediate that $k$ equivalence classes $[c_i]_\pi$ of $\pi$ (Eqn.~\ref{eqn:IIP_ec}) become $k$ equivalence index classes 
\begin{equation} \label{eqn:IP_eic}
[c_i]_{\pi^-} := \lbrace c_i, c_i+1, \ldots, c_{i+1}-1 \rbrace \text{ for } i=1,2,\ldots,k
\end{equation}
of the domain $[n]$ of $\pi^-$, with cardinalities $c'_1,c'_2,\ldots,c'_k$, respectively. For each canonical representative fixed index $c_i \in [c_i]_{\pi^-}$  of fixed element $a_i$, there are $(c'_i-1)$ idle indices $c_i+1, \ldots, c_{i+1}-1$ of the idle elements of $\pi$ equivalent to $a_i$ until the next canonical representative index $c_{i+1}$ of fixed element $a_{i+1}$. This means that, $k$ equivalence index classes $[a_i]_\pi$ of the domain $[n]$ of $\pi$ (Eqn.~\ref{eqn:IIP_ec_index}) become $k$ equivalence classes
\begin{equation} \label{eqn:IP_ec}
[a_i]_{\pi^-} := \lbrace  a_i, \pi^-(c_i+1), \ldots, \pi^-(c_{i+1}-1)  \rbrace \text{ for } i=1,2,\ldots,k
\end{equation}
of $\pi^-$, with cardinalities $c'_1,c'_2,\ldots,c'_k$, respectively. For each canonical representative fixed element $a_i \in [a_i]_{\pi^-}$, there are $(c'_i-1)$ idle elements $\pi^-(c_i+1), \ldots, \pi^-(c_{i+1}-1)$ equivalent to and coming immediately after the canonical representative fixed element $a_i$ located at $\pi^-(c_i)$ until the next canonical representative fixed element $a_{i+1}$ located at $\pi^-(c_{i+1})$. Hence, the elements of the equivalence classes are linearly ordered in $\pi^-$ such that $[a_1]_{\pi^-} < [a_2]_{\pi^-} < \ldots < [a_k]_{\pi^-}$. 

\end{cor}

It is easy to find out in $\pi^-$ the initial fixed and final idle elements of each equivalence class $[a_i]_{\pi^-}$, respectively, using the characteristic function $\varphi^-$, since inverting $\pi$ inverts the characteristic function as well, such that $\varphi^-(x)=1$ if $x \in C$, hence $\pi^-(x) \in A$, and $\varphi^-(x)=0$ otherwise, for all $x \in [n]$, and indicates the membership of an element in $C$ of $[n]$, as well as equivalently the membership of an element in $A$ of $\pi^-$. 

Since the first fixed element of $\pi$ is always $1$, the first fixed element of $\pi^-$ is always located at $\pi^-(1)$. Hence, the definition of inverse of an idempotent permutation becomes as follows. Inverse $\pi^-$ of an idempotent permutation $\pi$ of degree $k$ is a permutation of $[n]$, with $1 \le k \le n$ elements fixed and in increasing order with respect to each other starting strictly from $\pi(1)$. 

Although always will be generated by inverting an idempotent permutation, $\pi^-$ can be generated from $[n]$, as a result of following tasks:
\begin{enumerate}[label=(\arabic{*}).]
\item Select $k$ {\em elements} $A=\lbrace a_1,a_2,\ldots,a_k \rbrace$ from $[n]$ in $\binom{n}{k}$ ways.
\item Select strictly increasing $(k-1)$ {\em indices} $C=\lbrace 1,c_2,\ldots,c_k, n+1 \rbrace$ from $[n]$, in $\binom{n-1}{k-1}$ ways, and {\em fix} $k$ elements of $A$ in $\pi$ such that $\pi(c_i)=a_i$, for $i=1,2,\ldots,k$.
\item Arrange the remaining unfixed, hence {\em idle} $(n-k)$ {\em elements} of $\pi$ ($\complement{A}$) into $(n-k)$ empty locations of $\pi$ ($\complement{C}$)  in $\frac{(n-k)!}{(c'_1-1)!(c'_2-1)!\ldots (c'_k-1)!}$ different ways for the particular $C$ selected in the previous step.
\end{enumerate}

\begin{cor}
It should be noticed that, any algorithm that inverts $\pi$ for which $\varphi$ is the characteristic function, can be used to invert $\pi^-$ for which $\varphi^-$ is the characteristic function.
\end{cor}


\subsection{Surjection between $IP(n)$ and multisets}
Let $M(n)$ be the set of all multisets generated by {\em unordered} selection with replacement of $n$ elements from $[n]$, and $M(n,k)$ be the subset of those having $k$ distinct elements~\cite{rosen:handbook}.

\begin{thm}\label{thm:multiset}
For $1 \le k \le n$
\begin{equation}
\vert M(n) \vert = \sum^n_{k=1} \vert M(n,k) \vert = \sum^n_{k=1} \binom{n}{k} \binom{n-1}{k-1} 
\end{equation}
\end{thm}
\begin{proof} Each $m\in M(n,k)$ is a result of following two independent tasks: 
\begin{enumerate}[label=(\arabic{*}).]
\item Select $k$ distinct {\ elements} $A =\lbrace a_1,a_2,\ldots,a_k \rbrace$ of $m$ from $[n]$ in $\binom{n}{k}$ ways.
\item Select {\em nonzero} multiplicity $c'_i$ of each $a_i$ in $m$, for $i=1,2,\ldots,k$, such that $c'_1+c'_2+\ldots + c'_k=n$, in $\binom{n-1}{k-1}$ ways (the number of {\em nonzero} integral solutions). This is equivalent to selecting strictly increasing $(k-1)$ elements of $C=\lbrace 1,c_2,\ldots,c_k, n+1 \rbrace$ from $[n]$ in $\binom{n-1}{k-1}$ ways, where the difference sequence $c'_i = c_{i+1} - c_{i}$ is of length $k$ and $\sum^k_{i=1} c'_i=n$, since the fist and last elements are $1$ and $(n+1)$, respectively.
\end{enumerate}
Hence, the cardinality of $M(n,k)$ is $\binom{n}{k} \binom{n-1}{k-1}$ and that of $M(n)$ is $\sum^n_{k=1} \binom{n}{k} \binom{n-1}{k-1}$.
\end{proof}

Suppose that every multiset is linearly ordered. Thus, every $m \in M(n,k)$ can be defined by $m = \lbrace  c'_1 \cdot a_1,  c'_2 \cdot a_2, \ldots, c'_k \cdot a_k\rbrace$, where $c'_i$ denotes the multiplicity of $a_i \in m$, for $k=1,2,\ldots,k$.

\begin{thm}\label{thm:sur_IP_M_def}
There is a surjection from $IP(n)$ into $M(n)$.
\end{thm}

\begin{proof}
Let $\pi \in IP(n,k)$ and $m \in M(n,k)$, for which $A=\lbrace a_1,a_2,\ldots,a_k \rbrace$ and $C=\lbrace 1,c_2,\ldots,c_k,n+1 \rbrace$. Since $A$ and $C$ are common,  multiplicity of each $a_i \in m$ is equal to the cardinality of $[c_i]_\pi$. Hence, for particular $A$ and $C$, regardless of all possible arrangements of its idle elements, the inverse $\pi^-$ of $\pi \in IP(n,k)$ is equivalent to $m \in M(n,k)$ such that 
\begin{equation*}
\lbrace  [a_1]_{\pi^-}, [a_2]_{\pi^-}, \ldots, [a_k]_{\pi^-} \rbrace \sim  \lbrace c'_1 \cdot a_1,  c'_2 \cdot a_2, \ldots, c'_k \cdot a_k\rbrace
\end{equation*}
which implies that there is a surjection from $IP(n)$ into $M(n)$.
\end{proof}

\begin{thm}\label{thm:sur_IP_M}
The surjection from inverse $\pi^-$ of $\pi \in IP(n)$ into $m \in M(n)$ can be defined by the following ternary function $m=\mathcal{L}(\pi^-,\varphi^-,m)$.
\begin{equation}\label{eqn:sur_IP_M_ex}
m (i) = \begin{cases}
\pi^-(i) 	& \text{if  $\varphi^-(x)=1$},\\
m(i-1)	& \text{otherwise}.
\end{cases}
\quad \text{ for } i =1,2,\ldots,n
\end{equation}
\end{thm}
\begin{proof}

Let $\pi \in IP(n,k)$ and $m \in M(n,k)$, for which $A=\lbrace a_1,a_2,\ldots,a_k \rbrace$ and $C=\lbrace 1,c_2,\ldots,c_k,n+1 \rbrace$. Canonical representative fixed element $a_i$ of each $[a_i]_{\pi^-}$ corresponds to the first occurrence of $a_i \in m$. Hence, $m(c_i)=a_i$, for $i=1,2,\ldots,k$, which is equivalent to $m(i)=\pi^-(i)$ if $\varphi^-(i)=1$, for $i=1,2,\ldots,n$, since the characteristic function $\varphi^-$ identifies in $\pi^-$ the canonical representative fixed elements. On the other hand, the remaining idle elements of $[a_i]_{\pi^-}$ coming immediately after $a_i$ are equivalent to $a_i$ (Corollary~\ref{cor:inverse_I}). Hence, after $m(c_i)$ is set to $a_i$, it is evident that $m(c_i+1)=m(c_i)$, $m(c_i+2)=m(c_i+1)$, and this continues until the next canonical representative fixed element $a_{i+1}$, identified by $\varphi^-$, which implies  $m(i)=m(i-1)$ if $\varphi(i)=0$, for $i=1,2,\ldots,n$. Since $\pi^-(1)$ is always a fixed element from the definition of inverse of an idempotent permutation, the ternary function $\mathcal{L}(\pi^-, \varphi^-, m)$ defined in Eqn.~\ref{eqn:sur_IP_M_ex} determines $m$ from $\pi^-$. 
\end{proof}

\begin{cor}\label{cor:sur_IP_M_2}
The surjection from inverse $\pi^-$ of $\pi \in IP(n)$ into $m \in M(n)$ defined by Eqn.~\ref{eqn:sur_IP_M_ex} can be defined equivalently by the binary transformation $\pi^-=\mathcal{L}(\pi^-,\varphi^-)$ which transforms $\pi^-$ into $m$ such that,
\begin{equation}\label{eqn:sur_IP_M}
\pi^- (i) = \begin{cases}
\pi^-(i)  	& \text{if  $\varphi^-(i)=1$},\\
\pi^-(i-1)	& \text{otherwise}.
\end{cases}
\quad \text{ for } i =1,2,\ldots,n
\end{equation}

Furthermore, if the characteristic function $\varphi^-$ is implicitly defined in $\pi^-$, i.e., canonical representative fixed elements of $\pi^-$ are negative, then the above binary transformation becomes the unary transformation $\pi^-=\mathcal{L}(\pi^-)$ which transforms $\pi^-$ into $m$ such that,
\begin{equation}\label{eqn:sur_IP_M_unary}
\pi^- (i) = \begin{cases}
\lvert \pi^-(i) \rvert  	& \text{if  $\pi^-(i)<0$},\\
\pi^-(i-1)	& \text{otherwise}.
\end{cases}
\quad \text{ for } i =1,2,\ldots,n
\end{equation}
\end{cor}

\begin{thm} \label{thm:alg_IO_M}
There exists an $\mathcal{O}(n)$ time algorithm that, using $\log n$ bits, generates $m \in M(n)$ in-place of inverse $\pi^-$ of $\pi \in IP(n)$, where $\varphi^-$ is defined implicitly.
\end{thm}
\begin{proof}
Let $D[1\ldots n]$ be the array storing $\pi^-$, where fixed elements are negative. Using the unary transformation given in Eqn.\ref{eqn:sur_IP_M_unary}, following algorithm generates $m$ in-place of $\pi^-$ in $D[1 \ldots n]$, starting with $i \leftarrow 1$ and requires $\log n$ bits.
\begin{enumerate}[label=(\roman{*}).]
\item If $i>n$, then terminate. Otherwise, if $D_i < 0$, then set $D_i \leftarrow \vert D_i \vert$, increase $i$ by one and repeat this step; otherwise, set $D_i \leftarrow D_{i-1}$, increase $i$ by one and repeat this step.
\end{enumerate}
At the end, the array $D[1 \ldots n]$ stores $m = \lbrace c'_1 \cdot a_1,  c'_2 \cdot a_2, \ldots, c'_k \cdot a_k \rbrace$.
\end{proof}

\begin{thm}\label{thm:iterate_m_from_pi^-}
There exists an $\mathcal{O}(n)$ time algorithm that, using $2 \log n$ bits, iterates $m \in M(n)$ from inverse $\pi^-$ of $\pi \in IP(n)$, where $\varphi^-$ is defined implicitly.
\end{thm}
\begin{proof}
Let $D[1\ldots n]$ be the array storing $\pi^-$, where fixed elements are negative. Using the unary transformation given in Eqn.\ref{eqn:sur_IP_M_unary}, following algorithm iterates $m$ from $\pi^-$, starting with $i \leftarrow 1$ and requires $2 \log n$ bits.
\begin{enumerate}[label=(\roman{*}).]
\item If $i>n$, then terminate. Otherwise, if $D_i < 0$, then set $j \leftarrow \vert D_i \vert$, print $j$, increase $i$ by one and repeat this step; otherwise, print $j$, increase $i$ by one and repeat this step.
\end{enumerate}
\end{proof}


\subsection{Bijection between $IP(n)$ and $I(n)$}

\begin{thm} \label{thm:bijection_IP_I_def}
Let $\pi \in IP(n,k)$ and $\iota \in I(n,k)$, for which $A=\lbrace a_1,a_2,\ldots,a_k \rbrace$ and $C=\lbrace 1,c_2,\ldots,c_k,n+1 \rbrace$. If the equivalence index classes $[a_i]_\pi$ (Corollary~\ref{cor:ai_pi}) and $[a_i]_\iota$ (Corollary~\ref{cor:[a_i]}) are equal, then each $\pi(i)$ becomes the rank of $\iota(i)$ describing where to move it while rearranging the elements of $\iota$ in order with respect to each other. In other words, $\pi^-$ becomes the permutation of indices such that $\iota(\pi^-_1) < \iota(\pi^-_2) < \ldots < \iota(\pi^-_n)$.

\end{thm}
\begin{proof}
Since $A$ and $C$ are common, the cardinality of $I(n)$ (\thmref{thm:I}) is equal to that of $IP(n)$ (\thmref{thm:IP}). Furthermore, $\pi(a_i)=c_i$ and $\iota(a_i) = a_i$ imply that the fixed index $a_i$ of the fixed element $c_i$ of $\pi$ is equal to that of fixed element $a_i$ of $\iota$, for $i=1,2,\ldots,k$.  Hence, if $[a_i]_\pi=[a_i]_\iota$, for $i=1,2,\ldots,k$, then, when the idempotent permutation is inverted, the equivalence classes $[a_i]_{\pi^-}$ are linearly ordered in $\pi^-$ such that $[a_1]_{\pi^-} < [a_2]_{\pi^-} < \ldots < [a_k]_{\pi^-}$ (Corollary~\ref{cor:inverse_I}) and become equal to the equivalence index classes $[a_i]_\iota$ of $\iota$ which implies that $\pi^-$ becomes the permutation of indices such that $\iota(\pi^-_1) < \iota(\pi^-_2) < \ldots < \iota(\pi^-_n)$.
\end{proof}

\subsubsection{Obtaining $\iota \in I(n)$ from $\pi \in IP(n)$}

\begin{thm}\label{thm:iota_from_pi_okn}
There is an algorithm that generates $\iota \in I(n,k)$ in-place of $\pi \in IP(n,k)$ (for which $\varphi$ is defined implicitly) in $\mathcal{O}(k n)$ time using $4\log n$ bits.
\end{thm}
\begin{proof}
Since fixed element of an idempotent permutation is negative and in order with respect to the others, searching $\pi$ for $i=1,2,\ldots,n$, one can find the first fixed element $\pi(a_1)=-1$, and set to $a_1$. Then continue and find the second fixed element $\pi(a_2)=-c_2$, and set to $a_2$. Since the two fixed elements identify in $\pi$ the idle elements $2,3,\ldots, c_2-1$ (\thmref{thm:IP}), searching $\pi$ again for $j=1,2,\ldots,n$, these idle elements can be found and set to $a_1$, respectively. Afterwards, continuing the initial search on $i$, the third fixed element $\pi(a_3)=-c_3$ can be found and set to $a_3$ from where idle elements $c_2+1,c_2+2,\ldots, c_3-1$ of $\pi$ can be found and set to $a_2$, respectively, with another search for $j=1,2,\ldots,n$. When all the fixed elements of $\pi$ are processed in this way, $\pi$ becomes $\iota$ in $\mathcal{O}(k n)$ time.
\end{proof}

\begin{thm}\label{thm:bijection_IP_I}
Let $\pi^-$ be the inverse of $\pi \in IP(n,k)$, where $\varphi^-$ is the characteristic function. Then, $\iota \in I(n,k)$ can be determined from $\pi^-$ in $\mathcal{O}(n)$ time with the following ternary function $\iota=\mathcal{F}(\pi^-,\varphi^-,\iota)$,
\begin{equation}\label{eqn:bijection_IP_I}
\iota ( \pi^-(i) ) = \begin{cases}
\pi^-(i) 	& \text{if  $\varphi^-(i)=1$},\\
 \iota( \pi^-(i-1) ) 	& \text{otherwise}.
\end{cases}
\quad \text{ for } i =1,2,\ldots,n
\end{equation}

\end{thm}
\begin{proof}
$\iota(a_i) = a_i$ implies $\iota(\pi^-(c_i)) = \pi^-(c_i)$, for $i=1,2,\ldots,k$, which is equivalent to $\iota(\pi^-(i))=\pi^-(i)$ if $\varphi^-(i)=1$, for $i=1,2,\ldots,n$, since the characteristic function $\varphi^-$ identifies in $\pi^-$ the fixed elements. Since idle elements $\pi^-(c_i+1), \ldots, \pi^-(c_{i+1}-1)$ of equivalence class $[a_i]_{\pi^-}$ are all equivalent to and coming immediately after fixed element $\pi^-(c_i)=a_i$ (Corollary~\ref{cor:inverse_I}), then $\iota(\pi^-(c_i+1))=\iota(\pi^-(c_i))$, $\iota(\pi^-(c_i+2))=\iota(\pi^-(c_i+1))$, and this continues until the next fixed element $\pi^-(c_{i+1})$ identified by $\varphi^-$, which implies $\iota( \pi^-(i))=\iota( \pi^-(i-1))$ if $\varphi^-(i)=0$, for $i=1,2,\ldots,n$. Since $\pi^-(1)$ is always a fixed element from the definition of inverse of an idempotent permutation, the ternary function $\mathcal{F}(\pi^-, \varphi^-, \iota)$ defined in Eqn.~\ref{eqn:bijection_IP_I} uniquely determines $\iota \in I(n)$ from $\pi^-$ of $\pi \in IP(n)$.
\end{proof}

\begin{cor}\label{cor:bijection_IP_I_binary_T}
If the characteristic function $\varphi^-$ is defined implicitly in $\pi^-$, i.e., fixed elements of $\pi^-$ are negative, then, $\iota \in I(n,k)$ can be determined from $\pi^-$ in $\mathcal{O}(n)$ time with the following binary function $\iota=\mathcal{F}(\pi^-,\iota)$,
\begin{equation}\label{eqn:bijection_IP_I_binary_T}
\iota ( \vert \pi^-(i) \vert ) = \begin{cases}
\vert \pi^-(i) \vert 	& \text{if  $\pi^-(i) < 0$},\\
 \iota( \vert \pi^-(i-1) \vert ) 	& \text{otherwise}.
\end{cases}
\quad \text{ for } i =1,2,\ldots,n
\end{equation}
\end{cor}

\begin{thm}\label{thm:gen_iota_iip}
Corollary~\ref{cor:bijection_IP_I_binary_T} implies that there exists an $\mathcal{O}(n)$ time algorithm that, using $\log n$ bits, generates $\iota \in I(n)$ in a separate output array from $\pi^-$ where $\varphi^-$ is defined implicitly, i.e., fixed elements of $\pi^-$ are negative.
\end{thm}
\begin{proof}

Let $D[1\ldots n]$ be the array storing $\pi^-$, where fixed elements are negative. Using Eqn.\ref{eqn:bijection_IP_I_binary_T}, following algorithm generates $\iota$ in the output array $E[1 \ldots n]$ using $\log n$ bits, starting with $i \leftarrow 1$.
\begin{enumerate}[label=(\roman{*}).]
\item If $i>n$, then terminate. Otherwise, if $D_i < 0$, then set $E[ \: \vert D_i \vert \:] \leftarrow \vert D_i \vert$, increase $i$ by one and repeat this step; otherwise, set $E[ \: D_i \:] \leftarrow E[\vert \:D_{i-1}\vert \: ]$, increase $i$ by one and repeat this step.
\end{enumerate}
At the end, the array $E[1\ldots n]$ stores $\iota \in I(n)$. 
\end{proof}

\begin{thm}\label{thm:composition_mopi}
Let $\pi \in IP(n,k)$ and $m \in M(n,k)$ for which $A=\lbrace a_1,a_2,\ldots,a_k \rbrace$ and $C=\lbrace 1,c_2,\ldots,c_k,n+1 \rbrace$. Hence $m = \mathcal{L}(\pi^-,\varphi^-,m)$ (\thmref{thm:sur_IP_M}).  The result of the composition $m \circ \pi$ is always an idempotent map, i.e., $x = m \circ \pi$ implies $x \in I(n,k)$. 
\end{thm}
\begin{proof}
The image $x([n])$ is equal to the image $m([n])$. On the other hand, $m(c_i)=a_i$, (\thmref{thm:sur_IP_M}) and $\pi(a_i)=c_i$, for $i=1,2,\ldots,k$ (\thmref{thm:IP}). Hence, $x(a_i)=m(\pi(a_i))=m(c_i)=a_i$, which implies that the result $x$ of the composition $m \circ \pi$ is always an idempotent map since it has $k$ distinct and fixed elements $a_1,a_2,\ldots,a_k$, such that $x(a_i)=a_i$, for $i=1,2,\ldots,k$. 
\end{proof}

\begin{thm}
There is an $\mathcal{O}(n)$ time algorithm that, using $2 \log n$ bits, iterates $\iota \in I(n)$ in order of its elements from $\pi^-$, where $\varphi^-$ is defined implicitly, i.e., fixed elements are negative.
\end{thm}
\begin{proof}
\thmref{thm:composition_mopi} implies that $\iota = m \circ \pi$. Hence, $m$ is the linearly ordered arrangement of $\iota$, which can be iterated from $\pi^-$ in $\mathcal{O}(n)$ time using $2 \log n$ bits (\thmref{thm:iterate_m_from_pi^-}).
\end{proof}

\subsubsection{Obtaining $\pi \in IP(n)$ from $\iota \in I(n)$}

\begin{thm}\label{thm:pi_from_iota_okn}
There is an algorithm that generates $\pi \in IP(n,k)$ (where $\varphi$ is defined implicitly) in-place of $\iota \in I(n,k)$ in $\mathcal{O}(k n)$ time using $3\log n$ bits.
\end{thm}
\begin{proof}
In contrast to \thmref{thm:iota_from_pi_okn}, since fixed element of an idempotent map satisfies $\iota(x)=x$ and in order with respect to the others, searching $\iota$ for $i=1,2,\ldots,n$, one can find the first fixed element $\iota(a_1)=a_1$, and set to $-1$. Then another search for $j=1,2,\ldots,n$, idle elements of $\iota$ equal to $a_1$ can be found and set to $2,3,\ldots,c_2-1$, respectively. Then continuing the initial search on $i$, the second fixed element $\iota(a_2)=a_2$ can be found, and set to $-c_2$. Then another search for $j=1,2,\ldots,n$, idle elements of $\iota$ equal to $a_2$ can be found and set to $c_2+1,c_2+2,\ldots,c_3-1$, respectively. When $k$ fixed elements of $\iota$ are processed in this way, in $\mathcal{O}(k n)$ time, $\iota$ becomes $\pi$ for which $\varphi$ is defined implicitly, i.e., fixed elements of $\pi$ are negative.
\end{proof}

The following section describes the way to obtain $\pi$ in-place of $\iota$ in $\mathcal{O}(n)$ time using $\log n$ bits.

Since $\iota(x)=x$ if and only if $x \in A$, the characteristic function $\varphi: [n] \mapsto \lbrace 0,1 \rbrace$ of $\pi$ becomes the characteristic function of $\iota \in I(n)$ such that, $\varphi (x) = 1$ if  $\iota(x)=x$, and $\varphi (x) = 0$ otherwise, for all $x \in [n]$, indicating the membership of an element either in $A$ or $B$ of $[n]$, having the value $1$ for all elements of $A$ and the value $0$ for all elements of $B$. If $\varphi$ is defined implicitly in $\iota$ by making its fixed elements $a_1,a_2,\ldots,a_k$, negative, the definition of the characteristic function becomes $\iota(x) < 0$ if $x \in A$, and $x \in B$ otherwise, for all $x \in [n]$.

\begin{thm}
Either implicit or explicit, the characteristic function $\varphi$ defined on $\iota \in I(n)$ for the fixed elements that satisfy $\iota(x)=x$, lets to define recoverable partial functions on $\iota$ for which the domain of definition is $\iota(A)$.
\end{thm}
\begin{proof}
Suppose that the characteristic function $\varphi(x)$ is defined explicitly. Hence, $\varphi(x)=1$ if and only if $\iota(x)=x$, for all $x \in [n]$. In such a case, any partial function on $\iota$, for which the domain of definition is $\iota(A)$, affects only the fixed elements $\iota(a_1),\iota(a_2),\ldots,\iota(a_k)$, and can be recovered by the characteristic function with the following binary transformation.
\begin{equation}\label{eqn:g(a)=x}
\iota (x) = x \text{ if } \varphi(x) = 1, \text{ for all } x \in [n]
\end{equation}

On the other hand, if the characteristic function $\varphi(x)$ is defined implicitly in $\iota$ making fixed elements negative, then $\iota(x)<0$ if and only if $x\in [A]$, for all $x \in [n]$. In such a case, any partial function on $\iota$, for which the domain of definition is $\iota(A)$, affects only the fixed elements $\iota(a_1),\iota(a_2),\ldots,\iota(a_k)$, and can be recovered by the characteristic  function with the following unary transformation.
\begin{equation}\label{eqn:g(a)=x_implicit}
\iota (x) = x \text{ if } \iota(x) < 0, \text{ for all } x \in [n]
\end{equation}

\end{proof}

The image of the equivalence index class $[a_i]_\iota$ is $\iota([a_i]_\iota)=\lbrace a_i \rbrace$, for $i=1,2,\ldots,k$. Hence, assuming that the characteristic function $\varphi$ is explicitly defined on $\iota$, the following function 
\begin{equation} \label{eqn:g(a)=1}
\iota(x) = 1 \text{ if } \varphi(x) = 1, \text{ for all } x \in [n]
\end{equation}
affects only $\iota(A)$ making all $\iota(a_1),\iota(a_2),\ldots,\iota(a_k)$ equal to $1$. As a result, the image of equivalence index classes with multiplicity $c'_i > 1$ become $\iota([a_i]_\iota)=\lbrace 1, a_i \rbrace$, whereas those with multiplicity $c'_i = 1$ become $\iota([a_i]_\iota)=\lbrace 1\rbrace$, and it is always possible to recover original $\iota$ back by Eqn.\ref{eqn:g(a)=x}.

$\iota(B) \subset A$ implies $\iota(\iota(B)) \subset \iota(A)$. Hence, it is possible to define partial functions on $\iota$ for which the domain of definition is the subset of $\iota(A)$ covered by $\iota(\iota(B))$. For each $a_i \in A$, there are $(c'_i-1)$ indices $b \in B$ such that $\iota(b)=a_i$. Hence, after making all $ \iota(a_1),\iota(a_2),\ldots, \iota(a_k)$ equal to $1$ with the partial function defined by Eqn.\ref{eqn:g(a)=1}, one can continue with another partial function for which the domain of definition is the subset of $\iota(A)$ covered by $\iota(\iota(B))$ determined by $\varphi=0$, and count the multiplicity $c'_i$ of $a_i$ at $\iota(a_i)$ by
\begin{equation}
\iota(\iota(x)) = \iota(\iota(x))+1, \text{ if } \varphi(x) = 0, \text{ for all } x \in [n]
\end{equation}
As a result, the image of equivalence index classes with $c'_i > 1$ become $\iota([a_i]_\iota)=\lbrace c'_i, a_i \rbrace$, whereas those with $c'_i = 1$ become $\iota([a_i]_\iota)=\lbrace c'_i \rbrace$, and it is always possible to recover original $\iota$ back by Eqn.~\ref{eqn:g(a)=x}.

After counting the multiplicities, if a prefix sum is computed for all $x \in [n]$ over the fixed elements of $\iota$ identified by $\varphi$, the image of the equivalence index classes with $c'_i > 1$ become $\iota([a_i]_\iota)=\lbrace c_i+c'_i-1 ,a_i  \rbrace$, whereas those with $c'_i = 1$ become $\iota([a_i]_\iota)=\lbrace c_i+c'_i-1  \rbrace$ and it is always possible to recover original $\iota$ back by Eqn.~\ref{eqn:g(a)=x}.

Finally, with the last partial function on $\iota$ defined below, for which the domain of definition is again the subset of $\iota(A)$ covered by $\iota(\iota(B))$, every $\iota(b)$ obtains its value from its image $\iota(\iota(b))=\iota(a)$, which is initially equal to $(c_i+c'_i-1)$ and decrease it by one for the remaining. For each $\iota(b)$, first $\iota(\iota(b))$ should be decreased, because if $\iota(b)$ gets its value first, it cannot reach its image $\iota(\iota(b))$ anymore. Hence, first $\iota(b)$ should access $\iota(\iota(b))$ and decrease it by one. Then it can access $\iota(\iota(b))$ again and get the value one more than $\iota(\iota(b))$, as follows
\begin{equation}
\iota(\iota(x)) = \iota(\iota(x))-1 \text{ and } \iota(x) = \iota(\iota(x))+1 \text{ if } \varphi(x) = 0,  \text{ for all } x \in [n]
\end{equation}
As a result, the image $\iota([a_i]_\iota)$ of the equivalence index class $[a_i]_\iota$ become
\begin{equation}
\iota([a_i]_\iota) := \lbrace  c_i, c_i+1, \ldots, c_i+c'_i-1  \rbrace \text{ for } i=1,2,\ldots,k
\end{equation}
which is equal to the equivalence class $[c_i]_\pi$ (Corollary~\ref{cor:IIP_ec2}) and implies that $\iota \in I(n)$ is transformed into $\pi \in IP(n)$.

\begin{thm} \label{thm:I_2_IP_ip}
There is an $\mathcal{O}(n)$ time algorithm that computes $\pi \in IP(n)$ in-place of $\iota \in I(n)$ using $\log n$ bits of additional space.
\end{thm}
\begin{proof}
Let $D[1\ldots n]$ be the array storing $\iota \in I(n)$. If it is allowed to modify the elements of $\iota$ in the range $[-n,n]$, $\pi$ can be computed in-place of $\iota$ using $\log n$ bits, by the following algorithm.
\begin{enumerate}[label=(\roman{*}).]
\item If $D_i =i$, then $D_i\leftarrow -1$, for $i=1,2,\ldots,n$.
\item If $D_i > 0$, decrease $D[D_i]$ by one, for $i=1,2,\ldots, n$. 
\item Prefix sum negative elements of $D$, for $i=1,2,\ldots, n$. 
\item If $D_i > 0$, increase $D[D_i]$ by one and set $D_i \leftarrow -D[D_i] + 1$, for $i=n,n-1,\ldots, 1$.
\end{enumerate}
At the end, the array $D[1\ldots n]$ stores $\pi \in IP(n)$ for which $\varphi$ is defined implicitly. Furthermore, since the last step processes $\iota$ for $i=n,n-1,\ldots, 1$, the idle elements of $[c_i]_\pi$ are linearly ordered with respect to each other.
\end{proof}



\subsection{Associative permuting}

In-place inverting  $\pi \in I(n)$ and its inverse $\pi^-$ has already been introduced in Section~\ref{sec:inverse_IP}. Another operation that has a combinatorial interpretation on $\pi$ is in-place inverting fixed elements while permuting idle elements which will be called as associative permuting since it is a combination of permuting and inverting.



Inverting $\pi$ requires additional $n$ bits to tag each inverted element if the characteristic function $\varphi$ is defined implicitly. On the other hand, if only linearly ordered arrangement of $\iota \in I(n)$ is required, there is no need for additional $n$ bits. This is possible by in-place inverting fixed (hence negative) elements while permuting idle (hence positive) elements as will be described next.

Suppose that $\varphi$ is implicitly defined in $\pi$, i.e., fixed elements are negative. Similar to cycle leader permutation approach, starting with the first positive element, an outer cycle leader permutation can move only the positive elements to their final position ignoring the negative ones. This is possible since when a positive element $x \in \pi$ is moved to its final position $\pi(x)$, it will tag itself such that $\pi(x) = x$. If a positive element is moved onto a negative element, then until a positive element is encountered again, an inner cycle leader permutation can move only the negative elements to their final position storing negative of their former position, which is the same with inverting the negative elements. When a positive element is encountered again, the inner cycle leader permutation can stop and the outer cycle leader permutation can continue until all the positive elements are in-place permuted.

\begin{thm} \label{thm:I_ap}
There is an $\mathcal{O}(n)$ time algorithm that in-place inverts negative elements while permuting positive elements, using $3\log n$ bits of additional space. 
\end{thm}

\begin{proof}
First the algorithm will be given. Then the proof will follow.
\begin{enumerate}[label=(\roman{*}).]
\item Set $i \leftarrow 1$, and continue with next step.
\item $[$Outer cycle leader permutation$]$ If $i > n$, then terminate. Otherwise, if $\pi(i) < 0$ or $\pi(i) = i$, then increase $i$ by one and repeat this step; otherwise, set $j \leftarrow \pi(i)$, $\pi(i) \leftarrow \pi(j)$, $\pi(j) \leftarrow j$, and continue with next step.
\item If $\pi(i)$ is positive, then goto previous step; otherwise goto next step. 
\item $[$Inner cycle leader permutation$]$ Set $k \leftarrow \vert \pi(i) \vert$, $\pi(i) \leftarrow \pi(k)$, $\pi(k) \leftarrow -j$, $j \leftarrow k$, and goto previous step.
\end{enumerate}

If there are $1 \le k \le n$ negative elements, any negative element $\pi(x)$ is either a singleton cycle ($\pi(x) = -x$) or a part of another disjoint cycle. If it is a singleton cycle, then its inverse is equal to itself and there are not any positive or negative elements which will be moved to the place of $\pi(x)$. On the other hand, if it is a part of a disjoint cycle, the only case in which an inner cycle leader permutation can not be started on $\pi(x)$ is when there are not any positive elements involved in that particular disjoint cycle. Let two negative elements $\pi(x)$ and $\pi(y)$ form a disjoint cycle $(\pi_x\pi_y)$ without a positive element. This means $\pi(x)$ and $\pi(y)$ address each other such that $\pi(x) =-y$ and $\pi(y) =-x$. However, this contradicts with the definition of idempotent permutations that the fixed elements are in increasing order with respect to each other. In other words, if $\vert \pi(x) \vert < \vert \pi(y) \vert$, then $\vert \pi(x) \vert = y$ implies $\vert \pi(y) \vert  > y$ or $\vert \pi(y) \vert = x$ implies $\vert \pi(x) \vert  < x$. Therefore, there exists at least one positive element in every disjoint cycle which includes at least one negative element. On the other hand, if there are not any positive elements in $\pi$ or for $1 \le r < n$, all possible $r$-combinations of the positive elements form a disjoint cycle (singleton ones are indeed disjoint cycles), then there is only one arrangement for relatively ordering $k$ negative elements in remaining $k$ places, which implies that each negative element is indeed a singleton cycle and its inverse is equal to itself.
\end{proof}

After $\pi$ is associatively permuted, i.e., negative elements are inverted while positive elements are permuted, the resulting sequence $\gamma$ would be as follows in two line notation,
\begin{equation}
\gamma=
\begin{pmatrix}[c]
c_1   & 2  & \ldots & c_2-1 &  c_2 & c_2+1 & \ldots & c_3   & \ldots &  c_k & \ldots  &  n \\
-a_1  & 2  & \ldots & c_2-1 & -a_2 & c_2+1 & \ldots & -a_3  & \ldots & -a_k & \ldots  &  n 
\end{pmatrix}
\end{equation}

It is important to notice that, associatively permuting $\pi$ inverts $\varphi$ into $\varphi^-$, as well. On the other hand, if $\varphi$ is explicitly defined, then it should be rearranged while $\pi$ is associatively permuted.

\begin{thm}\label{thm:gen_ordered_iota_ap}
There is an $\mathcal{O}(n)$ time algorithm that generates linearly ordered arrangement of $\iota \in I(n)$ in-place of $\gamma$ ($\varphi$ is defined implicitly) using $\log n$ bits.
\end{thm}
\begin{proof} The algorithm defined in \thmref{thm:alg_IO_M} generates linearly ordered arrangement of $\iota \in I(n)$ in-place of $\gamma$.
\end{proof}

\section{Applications of idempotent permutations}\label{sec:application}

In this section, 3 different sorting applications will be introduced using idempotent permutations.

\begin{thm}
Given an array $D[1\ldots n]$ storing a map $f \in F(n)$, if it is allowed to modify the elements of $f$ in the range $[-n,n]$, then $f$ can be rearranged in-place in order of its elements unstably in $\mathcal{O}(n)$ time using only $4\log n$ bits in total. 
\end{thm}
\begin{proof} $f \in F(n)$ can be rearranged in-place in order of its elements by,

\begin{enumerate}[label=(\roman{*}).]
\item rearrange $f$ in-place into the idempotent map $\iota \in I(n)$ (\thmref{thm:F_2_I_ip}),
\item compute the idempotent permutation $\pi \in IP(n)$ in-place of $\iota$ (\thmref{thm:I_2_IP_ip}),
\item associatively permute $\pi$ into $\gamma$ (\thmref{thm:I_ap}),
\item generate linearly ordered arrangement of $\iota$ in-place of $\gamma$ (\thmref{thm:gen_ordered_iota_ap}),
\end{enumerate}
The rearrangement is unstable due to the fact that the idempotent map $\iota$ is obtained unstably from $f$ in the first task.
\end{proof}

\begin{thm}
Given an array $D[1\ldots n]$ storing a map $f \in F(n)$, if stability is important and it is not allowed to modify the elements of $f$ out of the range $[1,n]$, then $f$ can be rearranged in-place in order of its elements stably in $\mathcal{O}(n)$ time using an auxiliary array $E[1 \ldots n]$ plus $4\log n$ bits. 
\end{thm}
\begin{proof} $f \in F(n)$ can be rearranged in-place in order of its elements stably by,

\begin{enumerate}[label=(\roman{*}).]
\item compute $\sigma$ of $[n]$ in $E[1 \ldots n]$ from $f$ (\thmref{thm:F_2_I_sigma}),
\item rearrange $f$ stably into the idempotent map $\iota \in I(n)$ according to $\sigma$ (Corollary~\ref{rearranging_sigma}),
\item compute the idempotent permutation $\pi \in IP(n)$ in-place of $\iota$ (\thmref{thm:I_2_IP_ip}),
\item associatively permute $\pi$ into $\gamma$ (\thmref{thm:I_ap}),
\item generate linearly ordered arrangement of $\iota$ in-place of $\gamma$ (\thmref{thm:gen_ordered_iota_ap}),
\end{enumerate}
\end{proof}

\begin{thm}
Given an array $D[1\ldots n]$ storing a map $f \in F(n)$, if stability is important and it is not allowed to modify the elements of $f$, then $f$ can be rearranged in-place in order of its elements stably in $\mathcal{O}(n)$ time using an auxiliary array $E[1 \ldots n]$ plus $4\log n$ bits. 
\end{thm}
\begin{proof} $f \in F(n)$ can be rearranged in-place in order of its elements stably by,

\begin{enumerate}[label=(\roman{*}).]
\item compute $\sigma$ of $[n]$ in $E[1 \ldots n]$ from $f$ (\thmref{thm:F_2_I_sigma}),
\item rearrange $f$ stably into the idempotent map $\iota \in I(n)$ according to $\sigma$ (Corollary~\ref{rearranging_sigma}),
\item clear $E[1\ldots n]$ and compute the idempotent permutation $\pi \in IP(n)$ in $E[1\ldots n]$ from $\iota$ (\thmref{thm:I_2_IP_ip}),
\item rearrange $\iota$ in-place in order of its elements according to $\pi$ (Corollary~\ref{rearranging_sigma^-}).
\end{enumerate}
\end{proof}

\section{Conclusions}\label{chap:summaryandconclusion}

Idempotent permutations were introduced and their combinatorial interpretation was given. Together with a characteristic function defined either explicitly or implicitly, they uniquely represent the idempotent maps as well as their linearly ordered arrangement simultaneously. Moreover, in-place linear time transformations are possible between them. Hence, they make it possible to obtain linearly ordered arrangement of a map from $\lbrace 1,2,\ldots,n \rbrace$ into itself {\em in-place} in linear time using only $4 \log n$ bits in total, setting the theoretical lower bound of time and space complexity of sorting $n$ integer keys each in $[1,n]$. They may be important for other subjects as well, such as succinct data structures, information storing and searching.




\begin{thebibliography}{00}

\bibitem{knuth:vol1} {D.E. Knuth, The Art of Computer Programming, Volume 3: Sorting and Searching, Addison-Wesley, 1998.}
\bibitem{comtet}{L. Comtet, Advanced Combinatorics, Reidel, 1974}

\bibitem{Seward} {H.H. Seward, Information Sorting in the Application of Electronic Digital Computers to Business Operations, Master's thesis, MIT Digital Computer Laboratory, Report R-232, Cambridge, 1954.}

\bibitem{Feurzig} {W. Feurzig, ``Algorithm 23, mathsort'', Commun. ACM, Vol. 3, pp. 601~-~602, 1960.}


\bibitem{Isaac} {E.J. Isaac, R.C. Singleton, ``Sorting by address calculation'', J. of the ACM,  Vol. 3, pp. 169~-~174, 1956.}

\bibitem{Tarter} {M.E. Tarter, R.A. Kronmal, ``Non-uniform key distribution and address calculation sorting'', Proc. ACM Nat'l Conf. 21, 1966.}
\bibitem{Flores} {I. Flores, ``Computer time for address calculation sorting'', Journal of the ACM, Vol. 7, pp. 389~-~409, 1960.}
\bibitem{Jones} {B. Jones, ``A variation on sorting by address calculation'', Communications of the ACM , Vol. 13, pp. 105~-~107, 1970.}

\bibitem{Gupta} {G. Gupta, ``Sorting by hashing and inserting'', Proc. ACM Annual Computer Science Conf. 17, pp. 409~-~409, 1989.}
\bibitem{Suraweera} {F. Suraweera, J.M. Al-Anzy, ``Analysis of a modified address calculation sorting algorithm'', Comput. J. Vol. 31, pp. 561~-~563, 1988.}

\bibitem{mahmoud:2000} {H.M. Mahmoud, Sorting, A Distribution Theory, John Wiley and Sons, 2000.}
\bibitem{Cormen} {T.H. Cormen, C.E. Leiserson, R.L. Rivest, C. Stein, Introduction to Algorithms, The MIT Press, 2001.}

\bibitem{knuth:vol3} {D.E. Knuth, The Art of Computer Programming, Volume 1: Fundamental Algorithms, Addison-Wesley, 1997.}
\bibitem{Fich} {F.E. Fich, J.I. Munro, P.V. Poblete, ``Permuting in-place'', SIAM J. Comput., Vol. 24, pp. 266~–~278, 2006.}












\bibitem{rosen:handbook}{H.K. Rosen, Handbook of Discrete and Combinatorial Mathematics, CRC Press, 2000.}




\end{thebibliography}


\end{document}